\documentclass{aa}

\usepackage[varg]{txfonts}

\begin{document}

\titlerunning{Collisionless pair jet}
\authorrunning{M. E. Dieckmann et al.}

\title{Structure of a collisionless pair jet in a magnetized electron-proton plasma: flow-aligned magnetic field}

\author{M. E. Dieckmann\inst{1} \and D. Folini\inst{2} \and I. Hotz\inst{1} \and A. Nordman\inst{1} \and P. Dell'Acqua\inst{1} \and A. Ynnerman\inst{1} \and R. Walder\inst{2}}

\institute{Department of Science and Technology (ITN), Link\"oping University, 60174 Norrk\"oping, Sweden
\and 
\'Ecole Normale Sup\'erieure, Lyon, CRAL, UMR CNRS 5574, Universit\'e de Lyon, 69622 Lyon, France
}

\date{}

\abstract{}{We study the effect a guiding magnetic field has on the formation and structure of a pair jet that propagates through a collisionless electron-proton plasma at rest.}{We model with a particle-in-cell (PIC) simulation a pair cloud with the temperature 400 keV and mean speed 0.9c (c: light speed). Pair particles are continuously injected at the boundary. The cloud propagates through a spatially uniform, magnetized and cool ambient electron-proton plasma that is at rest. The mean velocity vector of the pair cloud is aligned with the uniform background magnetic field. The pair cloud has a lateral extent of a few ion skin depths.}{A jet forms in time. Its outer cocoon consists of jet-accelerated ambient plasma and is separated from the inner cocoon by an electromagnetic piston with a thickness that is comparable to the local thermal gyroradius of jet particles. The inner cocoon consists of pair plasma, which lost its directed flow energy while it swept out the background magnetic field and compressed it into the electromagnetic piston. A beam of electrons and positrons moves along the jet spine at its initial speed. Its electrons are slowed down and some positrons are accelerated as they cross the jet's head. The latter escape upstream along the magnetic field, which yields an excess of MeV positrons ahead of the jet. A filamentation instability between positrons and protons accelerates some of the protons, which were located behind the electromagnetic piston at the time it formed, to MeV energies.}{A microscopic pair jet in collisionless plasma has a structure that is similar to that predicted by a hydrodynamic model of relativistic astrophysical pair jets. It is a source of MeV positrons. An electromagnetic piston acts as the contact discontinuity between the inner and outer cocoons. It would form on subsecond time scales in a plasma with a density that is comparable to that of the interstellar medium in the rest frame of the latter. A supercritical fast magnetosonic shock will form between the pristine ambient plasma and the jet-accelerated one on a time scale that exceeds our simulation time by an order of magnitude.}


\maketitle

\section{Introduction}

Some X-ray binaries emit jets, which are composed of electrons, positrons and an unknown fraction of ions. Pair production by V404 Cygni during an outburst has been demonstrated by \citet{Siegert2016} and it is likely that some of these pairs enter the jet. Quantifying the baryon content of the jet is difficult because its radiation spectrum is dominated by synchrotron emissions of leptons \citep{Fender2014}. \citet{Trigo2013} found radio emissions from the jet of the X-ray binary 4U 1630-47 that are indicative of baryons but this observation could not be corroborated by \citet{Neilsen2014}. Only the X-ray binary SS433 is known to have a jet with a detectable baryon component \citep{Margon1979,Migliari2002,Waisberg2018}. Its jet is significantly slower than those of other X-ray binaries. \citet{Fender2000} remark that the energy budget available for accelerating the jet allows either for relativistic jets of electrons and positrons or nonrelativistic baryonic jets.

Some X-ray binaries with a black hole can emit at least intermittently jets, which expand at a relativistic speed into the surrounding medium \citep{Falcke1996,Mirabel1999,Fender2004,Siegert2016}. The latter can be interstellar medium (ISM) \citep{Ferriere2001} or stellar wind of the black hole's companion star. The ISM and stellar winds are an at least partially ionized dilute gas that consists mostly of hydrogen. Radiation from the accretion disk will ionize some of the gas in particular if it is an ultraluminous X-ray source \citep{Poutanen2007}. \citet{Waisberg2018} observed photo-ionization of material along the jet of SS433 by the radiation from its source. We can thus expect that relativistic jets of X-ray binaries interact with an ambient plasma with a significant number density.

\begin{figure}
\begin{center}
\includegraphics[width=0.8\columnwidth]{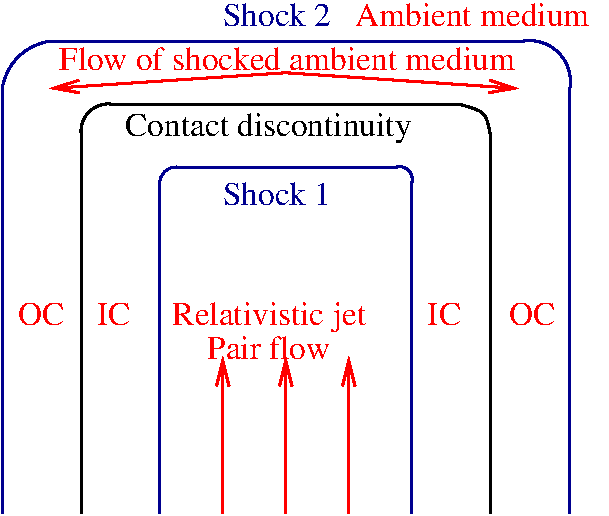}
\caption{Hydrodynamic jet model: a contact discontinuity separates the jet material from the outer cocoon, which is formed by the ambient material that was expelled by the jet. The outer cocoon is separated by the shock 2 from the ambient material that has not yet been affected by the jet. Shock 2 at the jet's head heats the ambient material that crosses the shock and lets it expand laterally. The shocked ambient plasma flows around the contact discontinuity and into the outer cocoon (OC). The energy required to displace the ambient material is provided by the jet material, which flows towards the contact discontinuity at a relativistic speed. It can not cross this discontinuity; it is slowed down and heated up as it approaches it. Shock 1 forms, which separates the fast-flowing jet material from the shocked jet material close to the contact discontinuity also known as the inner cocoon (IC). The thermal pressure of the inner cocoon pushes the discontinuity outwards.}\label{fig1}
\end{center}
\end{figure}

Interactions between the jet material and the surrounding ambient material have been studied with hydrodynamic models. A generally accepted hydrodynamic jet model (see \citet{Bromberg2011} and references therein) is that of a relativistic cylindrical jet with a planar head at its front that propagates into ambient material. It is discussed in Fig.~\ref{fig1}.  
Hydrodynamic models assume that binary collisions between particles occur fast enough to establish a thermal equilibrium of the material at any point on the relevant spatio-temporal scales. We can assign in this case unique bulk parameters such as mean speed, temperature and density to each point in the interacting material. Shocks and discontinuities enable rapid changes of these bulk parameters. 

Several X-ray binaries are close enough for us to resolve their jets in space and time. Their flow speed and temperature can thus be determined fairly accurately from experimental observations. Hydrodynamic simulations allow us to estimate the ratio between the mass density of the jet material and that of the ambient material. \citet{Massaglia1996} performed a parametric study, in which they varied the jet's flow speed and the density ratio. They compared the structure of the jets in their simulations to observed ones and they determined parameters for which both agree reasonably well. \citet{Massaglia1996} and \citet{DalPino2005} suggested that the mass density of the jet is a few per cent of that of the ambient material. 

Hydrodynamic models resolve well the macroscopic structure of jets, which is determined by global parameters such as the energy available to the jet and the resistance the ambient material offers to its expansion. The same is not always true for the microscopic structures. Hydrodynamic shocks between the ambient material and the outer cocoon and between the jet material and the inner cocoon and the contact discontinuity between both cocoons can only form in a collisional medium. 

\citet{Jean2009} estimate that a positron with an energy 1 MeV can propagate 30 kpc before it is decelerated by a gas with the number density 1 $\mathrm{cm}^{-3}$ to a speed at which it can recombine. Positrons with an energy that is comparable to those of the leptons carried by a mildly relativistic pair jet are thus stopped on a length scale that exceeds by far the size of the jet of an X-ray binary. Particle collisions may thus not always be sufficiently frequent to sustain hydrodynamic shocks and contact discontinuities, which involve particles of the jet and the ambient medium, in the relativistic jets of X-ray binaries. 

Structures in a plasma, in which effects due to binary collisions are negligible compared to those that involve the interplay of the plasma current with the electromagnetic field, can have similar properties as hydrodynamic shocks and contact discontinuities. They can form on time scales that are orders of magnitude smaller than collisional time scales. Such processes can be studied with particle-in-cell (PIC) codes. Their computational cost implies that we can only resolve jets that are microscopically small compared to that of an X-ray binary.   
 
Here we study with a PIC simulation how a pair cloud interacts with a magnetized ambient plasma. The magnetic field is aligned with the cloud's flow direction. The pair cloud swipes out the magnetic field in its way and compresses it into an electromagnetic piston that separates the ambient from the pair plasma. The separation is almost perfect at the sides of the pair cloud. Ambient protons are accelerated away from the jet by this electromagnetic piston and they reach a few percent of the speed of light $c$. The reflected protons form the outer cocoon. A magnetized collisionless shock  \citep{Shimada2000,Chapman2005,Caprioli2014,Lembege2018} will eventually form and separate the outer cocoon from the surrounding ambient plasma. The energy loss suffered by jet particles that interact with the electromagnetic piston slows them down and an inner cocoon forms. The inner cocoons on either side of the two-dimensional jet are separated by a beam of pairs that maintain their initial speed. The microscopic jet in our simulation thus has a structure that resembles that in Fig. \ref{fig1}.

Some positrons of the relativistic beam are accelerated as they cross the jet's head while its electrons are slowed down. The jet's head stripes off the pair beam's electrons and is thus a source of MeV positrons, which propagate along the magnetic field into the ISM. These positrons will eventually get stopped by their interaction with the particles and magnetic field of the ISM \citep{Jean2009,Panther2018}. 

Our paper is structured as follows. Section 2 discusses briefly the kinetic equations, on which collisionless plasma is based, and the numerical scheme of a PIC code. Relevant instabilities and previous work are summarized. The section concludes with listing our initial conditions. Section 3 shows the simulation results, which are summarized in Section 4. Section 4 also discusses future work addressing the matter content problem of jets.

\section{Collisionless plasma and previous work}

Infrequent binary collisions between particles imply that there is no constraint on the velocity distribution of particles of a given species $j$. Velocity becomes an independent variable and the ensemble of all particles of species $j$ is described by a phase space density distribution $f_j(\mathbf{x},\mathbf{v},t)$. It describes the probability, with which we find a particle at the position $\mathbf{x}$ with the velocity $\mathbf{v}$ at the time $t$. Each species $j$ of a collisionless plasma can be described by such a distribution. 

A particle species is in a thermal equilibrium if its density is uniform in space and if its velocity distribution is a nonrelativistic Maxwellian or a relativistic Maxwell-J\"uttner distribution that is isotropic in velocity space. The thermal velocity spread is defined in this case as $v_{th,j}={(k_B T_j/m_j)}^{(1/2)}$ ($k_B, T_j, m_j$: Boltzmann constant, temperature and particle mass). 

Individual plasma particles are charged and their microscopic current density is proportional to their velocity. By summing up the charge- and current density distributions of all particles of species $j$ we obtain its charge density $\rho_j (\mathbf{x},t) = \int f_j(\mathbf{x},\mathbf{v},t) \, d\mathbf{v}$ and the macroscopic current density $\mathbf{J}_j (\mathbf{x},t) = \int \mathbf{v} \, f_j(\mathbf{x},\mathbf{v},t) \, d\mathbf{v}$. A summation over all species yields the total charge density $\rho = \sum_j \rho_j$ and current density $\mathbf{J}=\sum_j \mathbf{J}_j$. Both are coupled to the macroscopic electric $\mathbf{E}$ and magnetic $\mathbf{B}$ fields via the Maxwell equations
\begin{equation}
\mu_0 \epsilon_0 \frac{\partial}{\partial t}\mathbf{E} = \nabla \times \mathbf{B} - \mu_0 \mathbf{J}, \label{Ampere}
\end{equation}
\begin{equation}
\frac{\partial}{\partial t}\mathbf{B}=-\nabla \times \mathbf{E}, \label{Faraday}
\end{equation}
\begin{equation}
\nabla \cdot \mathbf{E}=\rho/\epsilon_0, \nabla \cdot \mathbf{B}=0.\label{others}
\end{equation}
$\epsilon_0,\mu_0$ are the vacuum permittivity and permeability. Both field components act back on the particle $i$ of species $j$ via the relativistic Lorentz force
\begin{equation} 
\frac{d}{dt}\mathbf{p}_i = q_j \left ( \mathbf{E} + \mathbf{v} \times \mathbf{B} \right ).
\end{equation}
where $\mathbf{p}_i = m_j \Gamma_i \mathbf{v}_i$ and $\Gamma_i^{-2}=(1-\mathbf{v}_i^2/c^2)$.

Absent collisions between particles allow for many features not found in a collisional medium. Several beams of charged particles with different mean speeds can, for example, coexist in the same spatial interval. These beams relax via the electromagnetic fields that are driven by collisionless plasma instabilities. The filamentation instability of counterstreaming beams of charged particles (a review is provided by \citet{Bret2010}) or the \citet{Weibel1959} instability of one plasma species with a thermal anisotropy were examined with PIC simulations in order to determine if they can generate magnetic fields and shocks in electron-positron plasma \citep{Kazimura1998,Silva2003,Hededal2005,Chang2008,Dieckmann2018,Plotnikov2018} and in electron-ion plasma \citep{Spitkovsky2008}. 

\citet{Amato2006} investigated pair plasma with a minor fraction of ions for the case of ultrarelativistic shocks with a strong transverse magnetic field. They found that the pairs form increasingly nonthermal distributions as the fraction of ions in the upstream medium is increased and that the heating is stronger for positrons than for electrons. However, the one-dimensional geometry of their simulation suppressed the filamentation instability, which is dominant for such high shock speeds. 

The interaction of a hot expanding pair cloud with a cooler unmagnetized ambient electron-proton plasma was studied by \citet{Dieckmann2018a} in one spatial dimension. They found that the streaming electrons and positrons drive ion acoustic solitary waves. These waves eventually break and form electrostatic shocks \citep{Malkov2016}, which can accelerate protons to high energies. \citet{Dieckmann2018b} studied this expansion in two dimensions. A filamentation instability between the cloud particles and the ambient plasma resulted in the growth of magnetic fields that separated in space the protons and positrons.

All aforementioned simulations assumed that the plasma is uniform orthogonal to the plasma flow direction and shock normal. Collisions of cylindrical plasma clouds with a spatially uniform plasma have also been modeled. \citet{Nishikawa2016} examined how cylindrical clouds of electrons and positrons or electrons and ions interact with an electron-ion plasma. A three dimensional geometry was resolved and a reduced ion mass was used to speed up the simulation. The cloud particles had a low temperature and a highly relativistic mean speed. They interacted with the ambient plasma via Weibel and mushroom instabilities \citep{Alves2015}. \citet{Nishikawa2017} investigated the effects of a helical magnetic field for a similar plasma configuration. \citet{Arxiv} considered a hot mildly relativistic pair cloud that expanded into an unmagnetized electron-proton plasma. A jet formed. Magnetic fields due to the Weibel instability separated the jet plasma from the ambient plasma and acted as a contact discontinuity in a collisionless plasma. 

\section{Code and initial conditions}

The PIC simulation code EPOCH defines $\mathbf{E}$ and $\mathbf{B}$ on a numerical grid and updates them in time with a numerical approximation of Eqns. \ref{Ampere} and \ref{Faraday}. The current density $\mathbf{J}$ is obtained from the plasma. Each species $j$ is approximated by an ensemble of computational particles (CPs), which have the same charge-to-mass ratio as the species $j$. The electromagnetic fields are interpolated from the grid to the position of each CP and its momentum is updated with a numerical approximation of the relativistic Lorentz force. Each CP carries with it a current that is deposited on the grid using the \citet{Esirkepov2001} scheme. The summation of the contributions to the current density of all particles of all species $j$ yields $\mathbf{J}$, which is then used to update the electromagnetic fields. EPOCH fulfills Gauss' law and $\nabla \cdot \mathbf{B}=0$ to round-off precision and is discussed in detail by \citet{Arber2015}. 

We resolve the x-y plane and all three velocity components of the CPs. The simulation's box length along $x$ is $L_x$ and $0 \le x \le L_x$. The values of $y$ span the interval $-L_y/2 \le y \le L_y/2$. The boundary conditions are periodic along $y$ and reflecting along $x$. The simulation box is filled with a spatially uniform ambient plasma, which consists of electrons and protons with identical number densities $n_0$ and temperatures $T_0$ = 2 keV. All plasma density distributions will be normalized to $n_0$. Each species is represented by 14 CPs per cell. The plasma frequencies of the electrons with the mass $m_e$ and protons with the mass $m_p=1836m_e$ are $\omega_{pe} = {(n_0e^2/\epsilon_0m_e)}^{1/2}$ and $\omega_{pp}=\omega_{pe}\sqrt{m_e/m_p}$, respectively ($e$: elementary charge). 

Space is normalized by the proton skin depth $\lambda_s = c/\omega_{pp}$ as $x\rightarrow x/\lambda_s$. Our simulation grid resolves $L_x=24.6$ with 14000 cells and $L_y = 12.3$ with 7000 cells. The ambient plasma is permeated at the time $t=0$ by a magnetic field $\mathbf{B}_0=(B_0,0,0)$ with an amplitude that yields the electron gyro-frequency $\omega_{ce} = eB_0/m_e=\omega_{pe}/11.3$. The magnetic energy density or pressure $P_{B0}=B_0^2/2\mu_0$ equals the thermal pressure $P_{TE}=n_0k_B T_0$ of the electrons ($k_B$: Boltzmann constant). We express $\mathbf{E}$ and $\mathbf{B}$ in units of $c\omega_{pe}m_e/e$ and $\omega_{pe}m_e/e$.

The pair cloud, which will drive a jet into the ambient plasma in our simulation, consists of electrons and positrons with the temperature 400 keV. This value is comparable to the brightness temperature 100 keV given by \citet{Dhawan2000} for the jet of GRS 1915+105. The brightness temperature can serve as an estimate for the average particle temperature within the jet. 

\begin{figure}
\includegraphics[width=\columnwidth]{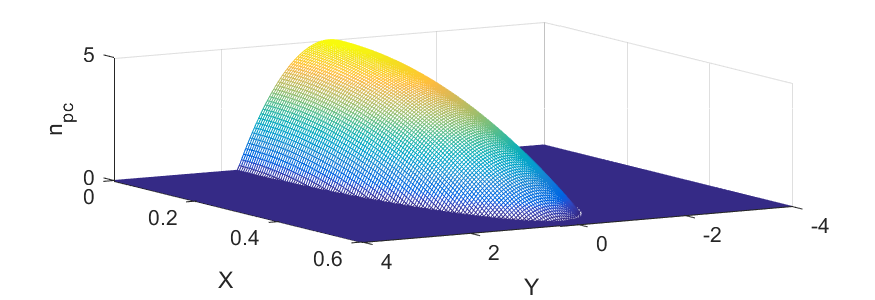}
\caption{The initial density distribution $n_{pc}=5-(45x^2+5y^2)/W_{jet}^2$ with $W_{jet}=1.7$ of the cloud's positrons and electrons.}\label{fig2}
\end{figure}

We select the mean speed 0.9c for the jet plasma, which is comparable to the speed of the superluminal jet of GRS 1915+105. A ratio of 0.005 is taken between the mass density of the pair cloud on its spine and that of the ambient plasma. This ratio is similar to the value proposed by \citet{Massaglia1996}. The density distribution of our pair cloud is $n_{pc}(x,y)=5-(45x^2+5y^2)/W_{jet}^2$ if $n_{pc}(x,y) \ge 0$ and zero otherwise (see Fig. \ref{fig2}).
The half-width of the jet is $W_{jet}=1.7$.

Each cloud species is resolved by $2\cdot 10^8$ CPs at the time $t=0$. Additional pairs with the density distribution $n_{jet}(y)=5-5y^2/W_{jet}^2$ for $|y|<W_{jet}$ are injected next to the boundary at $x=0$. They maintain the density distribution at the cross section $x=0$ of the cloud distribution in Fig. \ref{fig2} while the cloud is expanding to larger $x$. We inject $1.6 \cdot 10^5$ CPs at every time step. Time is normalized by $\omega_{pp}^{-1}$ as $t\rightarrow t\omega_{pp}$. The simulation time $t_{sim}=34$ is resolved by 62500 steps. 

The dynamics of a collisionless plasma that obeys the Maxwell-Lorentz set of equations does not change qualitatively with the value of $n_0$ as long as we keep the density ratios of all plasma species and the ratio between the electron plasma- and gyrofrequency unchanged. Space, time and $B_0$ scale in this case with $\lambda_s$, $\omega_{pp}^{-1}$ and $\omega_{pp}$. Table \ref{table} gives the physical values of $t_{sim}, W_{jet}$ and $B_0$ for several densities $n_0$ of the ambient plasma. 

\begin{table}[ht]
\begin{tabular}{|c|c|c|c|}
\hline
$n_0$ in $cm^{-3}$ & $t_{sim}$ in ms & $W_{jet}$ in $10^6$ m & $B_0$ in nT \\
\hline
0.001 & 820 & 12.3  & 0.9 \\ 
1       &   26 & 0.39  & 28 \\
100   &  2.6 & 0.039 & 280 \\
\hline
\end{tabular}
\caption{Physical values for the simulation time $t_{sim}$, the jet's half-width $W_{jet}$ and the initial amplitude $B_0$ of the background magnetic field for selected values of the ambient plasma density $n_0$.}
\label{table}
\end{table}

\section{Simulation results}

We discuss in the first subsection the global evolution of the particle and field distributions at the times $t_1=6.8$, $t_2=13.6$, $t_3=20.4$, $t_4=27.3$ and $t_{sim}$. Effects due to the proton gyromotion can be neglected because $\omega_{cp}/\omega_{pp} \approx 0.002$ ($\omega_{cp}=\omega_{ce}/1836$: proton gyro-frequency). We observe the formation and early evolution of the external shock. The second subsection examines the plasma close to the external shock at the time $t_{sim}$.

Electromagnetic waves emitted at $x=0$ at the time $t=0$ travel to the boundary at $x=24.6$, are reflected by it and return to $x=15.2$ at the time $t_{sim}$. Processes in the interval $0 \le x \le 14$ are thus not affected by the reflecting boundary at $x=24.6$ for times $t\le t_{sim}$ and we limit our investigation to this spatial window. 

\subsection{Global evolution}

\begin{figure}
\includegraphics[width=\columnwidth]{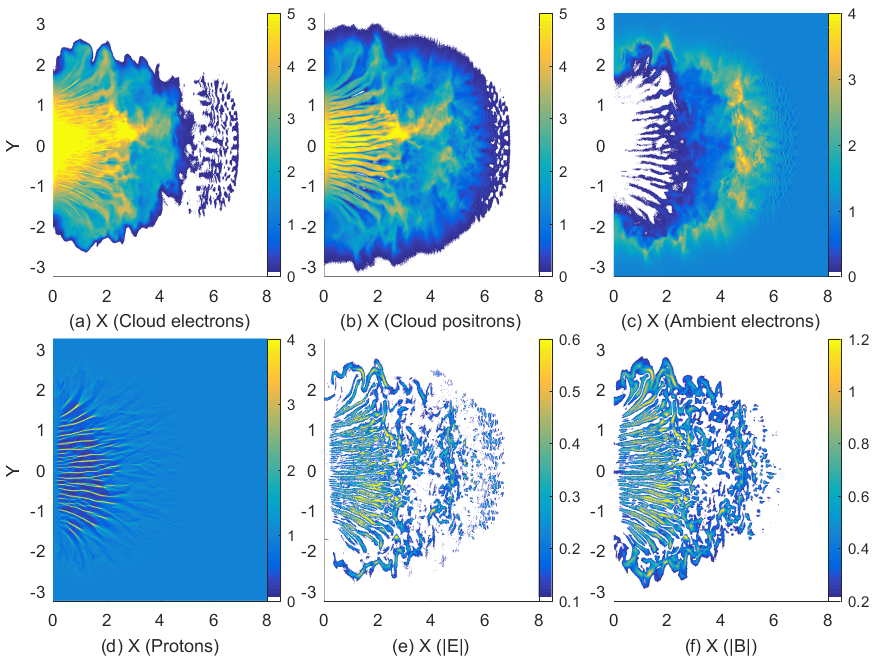}
\caption{The plasma and field distributions at $t_1=6.8$: panels (a) and (b) show the distributions of the cloud electrons and positrons. The electrons and protons of the ambient plasma are shown in panels (c) and (d), respectively. Panel (e) shows the electric field modulus $|\mathbf{E}|$ and $|\mathbf{B}|$ is shown in (f). The electric and magnetic amplitudes are clamped to values 0.1 and 0.2 to remove the noise.}\label{fig3}
\end{figure}

Unless stated otherwise all data was averaged over 4 cells along x and over 4 cells along y. Figures \ref{fig3}(a, b) show the density distributions of the cloud's electrons and positrons at $t=t_1$. Their density peaks at $x=0$ in the interval $|y|\le 1.7$ where they are injected at the left boundary. 
Both distributions extend well beyond their initial front in Fig. \ref{fig2}. The bulk of the cloud electrons is confined to values $x\le 5$ and $|y|\le 2.5$ while the front of the positrons has propagated farther by the distance $\approx$ 1. The front of the pair cloud has crossed the distance $\approx$ 5 along $x$ during the time interval $t_1$. Its speed is thus $\approx 0.7 c$. It has crossed the distance $\approx$ 0.8 along $y$, giving the lateral expansion speed $0.12c$. 

The distance from the initial cloud border $W_{jet}$ to $y=2.5$ is comparable to the normalized relativistic gyro-radius $r_{ge} = \Gamma_0 m_e V_0/(eB_0\lambda_s) \approx 0.54$ ($\Gamma_0 = 1/{(1-V_0^2/c^2)}^{1/2}$) of a lepton moving with the speed $V_0$. The rapid initial expansion of the pair cloud along $y$ and across the magnetic field is caused by a gyromotion of its particles. Dilute electron and positron populations, which originated from the front of the initial cloud distribution at $x\approx 0.5$, have reached $x\approx 7$. Their speed is just below $c$. 

The ambient electrons in Fig. \ref{fig3}(c) were evacuated at low values of $x,|y|$ in the spatial region that is occupied by cloud electrons. They have been compressed in the interval, in which the front of the positrons is located. Figures \ref{fig3}(a-c) thus demonstrate that the cloud's electrons are slowed down by their interaction with the ambient plasma while the positrons are accelerated.
 
\citet{Dieckmann2018} examined the expansion of a shock-heated pair cloud into a cooler unmagnetized pair plasma at a speed $\approx V_0$. The shock-heated pair cloud interacted with the ambient pair plasma via the two-stream instability, which saturated by forming electrostatic phase space vortices in the electron and positron distributions. An electron phase space vortex is tied to a local excess of positive charges. Trapped electrons gyrate in the associated positive potential. A localized negative potential traps positrons. Certain combinations of the electric field and the particle distributions yield stable nonlinear structures \citep{Schamel1986}. The symmetry between electrons and positrons and thus between their nonlinear structures resulted in an equal expansion speed of both species. 

In the case we consider here the electrons of the cloud can interact with those of the ambient plasma by forming phase space vortices. A phase space vortex mixes them in phase space. This mixing transports the ambient electrons away from the cloud's source in the propagation direction of the phase space hole. Positron phase space vortices can not trap and accelerate protons due to their large mass difference (see the simulation by \citet{Dieckmann2018a} for a detailed discussion). The population of the comoving ambient and cloud electrons is thus denser than that of the positrons. Any excess of negative current drives a positive electric field via Amp\`ere's law, which decelerates the electrons and accelerates the positrons. The fastest positrons eventually outrun the slower mixed electron population and build up a layer ahead of them that carries an excess positive current. This current drives an electric field, which accelerates and compresses the ambient electrons found in the layer with a density $2-4$ in Fig. \ref{fig3}(c) close to the pristine ambient electrons. 

\begin{figure}[ht]
\includegraphics[width=\columnwidth]{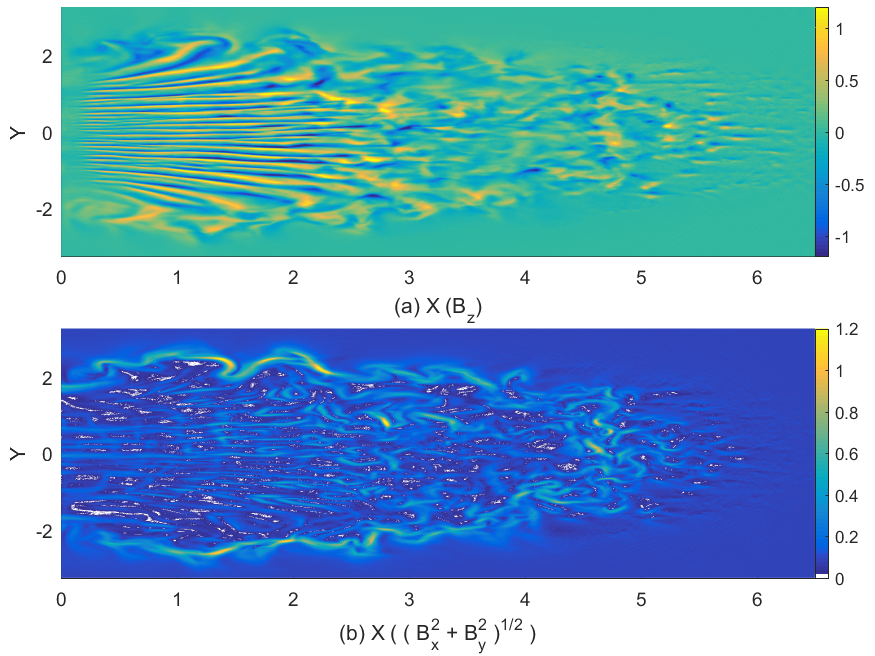}
\caption{The magnetic field distributions at $t_1=6.8$: panel (a) shows the out-of-plane magnetic field $B_z$, which is driven by the filamentation instability. Panel (b) shows the modulus of the in-plane magnetic field.}
\label{fig4}
\end{figure}

Figure \ref{fig4} compares the magnetic $B_z$ component, which is driven by the filamentation instability in the considered geometry, and the modulus of the in-plane magnetic field. 
We observe peak values of the magnetic field that exceed $B_0$ by an order of magnitude. The in-plane magnetic field in Fig. \ref{fig4}(b) is weaker than $B_z$ at this time except in some locations close to the boundary of the pair cloud.
The supplementary movie 1 animates Fig. \ref{fig4} in time interval $0 \le t \le t_{sim}$. It shows how the $B_z$ component driven by the filamentation instability weakens in time while the magnetic band in the in-plane magnetic field distribution strengthens in time.

The protons close to $x=0$ in Fig. \ref{fig3}(d) have been compressed into thin filaments. The pair cloud and the ambient plasma interact via a filamentation instability, which is responsible for the strong magnetic field in Figs. \ref{fig4}(a) that follows the proton density filaments.  The electromagnetic fields close to the cloud front in Fig. \ref{fig3} have a different structure and they are a result of the aforementioned two-stream instability, which is not purely electrostatic for relativistic speeds \citep{Bret2010}.  

\begin{figure*}
\includegraphics[width=\textwidth]{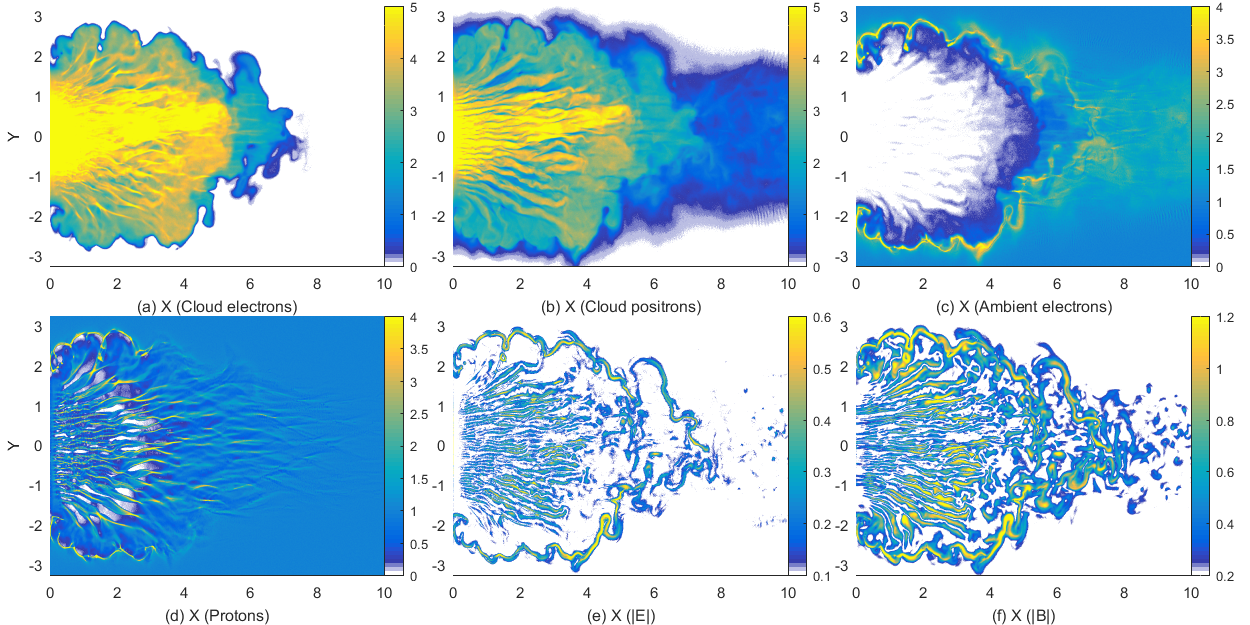}
\caption{The plasma and field distributions at $t_2=13.6$: panels (a) and (b) show the distributions of the cloud electrons and positrons. The electrons and protons of the ambient plasma are shown in panels (c) and (d), respectively. Panel (e) shows the electric field modulus $|\mathbf{E}|$ and $|\mathbf{B}|$ is shown in (f). The electric and magnetic amplitudes are clamped to values 0.1 and 0.2 to remove noise.}\label{fig5}
\end{figure*}

Figure \ref{fig5} shows the plasma and field distributions at the time $t_2$. The cloud front at $x\approx 2$ has expanded from $y\approx 2.6$ at $t=t_1$ to $y\approx 2.9$ at $t=t_2$, which yields a front speed of about $0.04c$ along $y$; the lateral expansion of the cloud has been slowed down by $B_0$. Figures \ref{fig3}(a) and \ref{fig5}(a) show that the cloud electrons have expanded from $x\approx 5$ at $t_1$ until $x\approx 7$ at $t_2$, which yields the speed $c/3$. 
Positrons have expanded farther in all directions and a beam with a density $\sim 0.5$ has formed at large $x$ and $|y|\le 2$. 

The pair cloud continued to expel ambient electrons in Fig. \ref{fig5}(c). A thin band with a density value $\approx 4$ marks in Fig. \ref{fig5}(c) the end of the spatial interval, which is occupied by the cloud plasma. A similar high-density band is seen in the protons for $x\le 4$ and $|y| \approx 2.8$. The high-density band is accompanied by strong electromagnetic fields in Fig. \ref{fig5}(e,f). The latter also show strong filaments at low values of $x,|y|$, which follow those in the proton density distribution in Fig. \ref{fig5}(d). The filamentation of the protons has continued and their density has decreased to almost zero in extended spatial intervals. 

\begin{figure*}
\includegraphics[width=\textwidth]{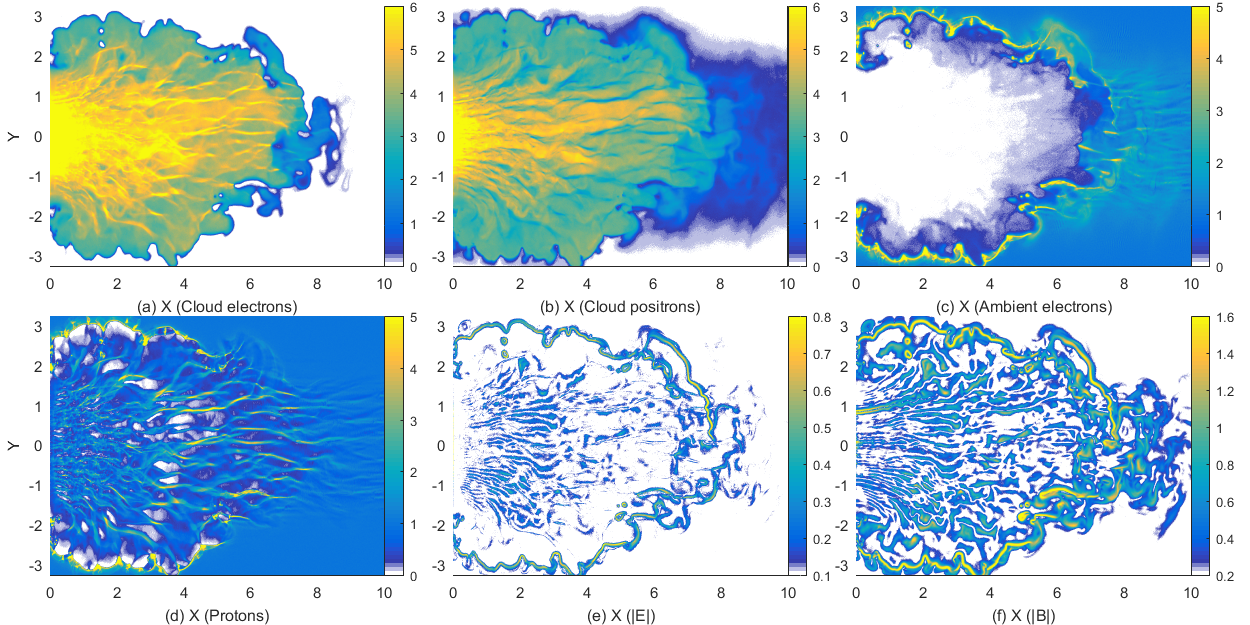}
\caption{The plasma and field distributions at $t_3=20$: panels (a) and (b) show the distributions of the cloud electrons and positrons. The electrons and protons of the ambient plasma are shown in panels (c) and (d), respectively. Panel (e) shows the electric field modulus $|\mathbf{E}|$ and $|\mathbf{B}|$ is shown in (f). The electric and magnetic amplitudes are clamped to values 0.1 and 0.2 to remove noise.}\label{fig6}
\end{figure*}

The cloud electrons at the time $t_3$ are confined to a smaller spatial interval in Fig. \ref{fig6}(a) than the positrons in Fig. \ref{fig6}(b). 
Only electrons from the cloud can compensate the charge imbalance caused the expulsion of the ambient electrons in Fig. \ref{fig6}(c). Hence they accumulate where the ambient electrons were expelled. 

The ambient electrons and protons still show a high density band. It was located at $|y|\approx 2.8$ at $t=t_2$ and $x\approx 2$ and at $|y| \approx 3.1$ at the same value of $x$ in Fig. \ref{fig6}(d) and the expansion speed along $y$ is $\approx 0.04c$. No high-density band can be observed at the front of the cloud at $x\approx 7$ at low $|y|$. Positrons flow out along $B_0$ in this y-interval and their charge is compensated by an accumulation of the ambient electrons. A filamentation instability continues to develop in the interval $x<4.5$ and $|y| \le 2$ between the particles of the pair cloud and the protons and we observe the strongest proton density filaments at $x\approx 4$ and $|y| \le 1.5$. The electromagnetic fields associated with the high-density band that encloses the pair cloud are now stronger than those driven by the filamentation instability within the pair cloud.

\begin{figure*}
\includegraphics[width=\textwidth]{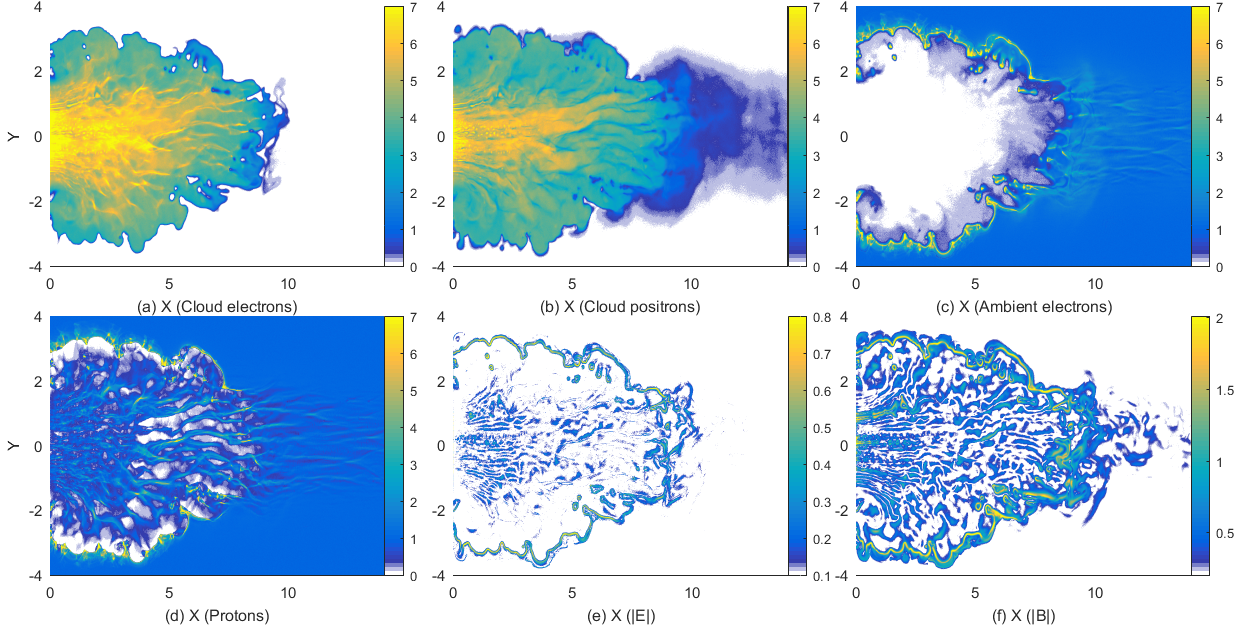}
\caption{The plasma and field distributions at $t_4=27.5$: panels (a) and (b) show the distributions of the cloud electrons and positrons. The electrons and protons of the ambient plasma are shown in panels (c) and (d), respectively. Panel (e) shows the electric field modulus $|\mathbf{E}|$ and $|\mathbf{B}|$ is shown in (f). The electric and magnetic amplitudes are clamped to values 0.1 and 0.2 to remove noise.}\label{fig7}
\end{figure*}

Figure \ref{fig7} shows that the high-density band and its associated electromagnetic fields mark the boundary between the cloud in Fig. \ref{fig7}(a,b) and the ambient electrons and protons in Figs. \ref{fig7}(c,d).  
The protons have been completely expelled from an interval with the width $\sim 0.25$ directly behind the high-density band. Figures \ref{fig7}(e,f) reveal a continuing weakening of the electromagnetic fields, which are associated with the filamentation instability between the pair cloud and the protons in the interval $x\le 4$ and $|y|\le 2$ while those associated with the high-density band have maintained their strength.

\begin{figure*}
\includegraphics[width=\textwidth]{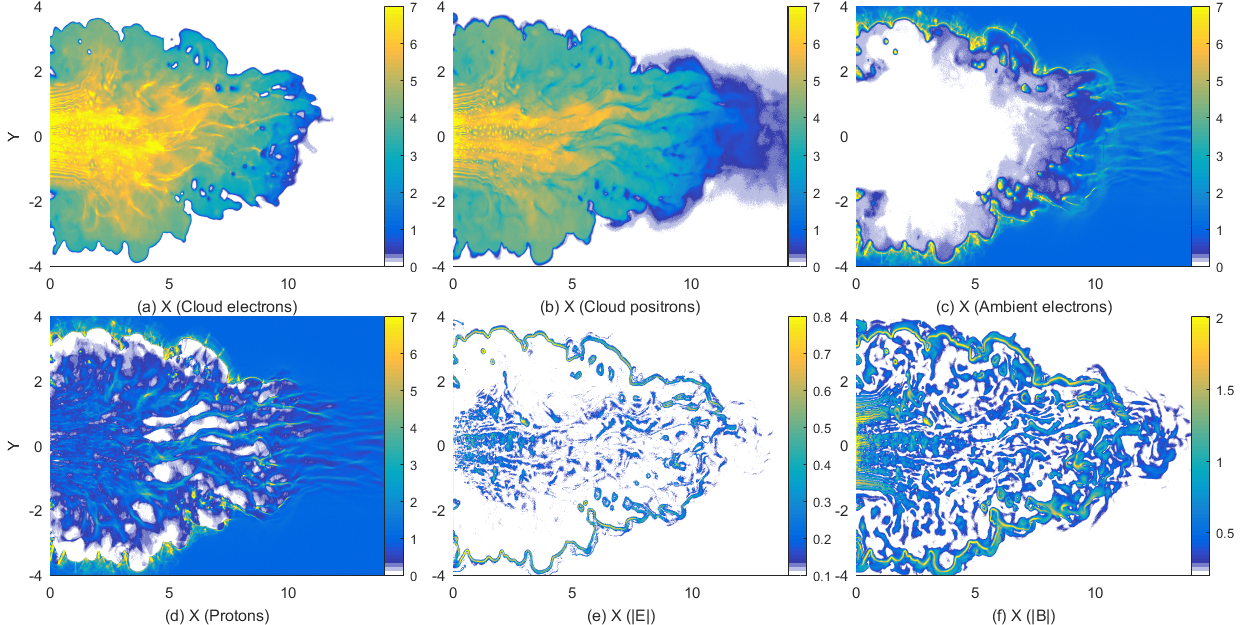}
\caption{The plasma and field distributions at $t_{sim}=34$: panels (a) and (b) show the distributions of the cloud electrons and positrons. The electrons and protons of the ambient plasma are shown in panels (c) and (d), respectively. Panel (e) shows the electric field modulus $|\mathbf{E}|$ and $|\mathbf{B}|$ is shown in (f). The electric and magnetic amplitudes are clamped to values 0.1 and 0.2 to remove noise.}\label{fig8}
\end{figure*}

Figure \ref{fig8} shows the distributions of the plasma species and of the electromagnetic fields at the final time $t_{sim}$. Figure \ref{fig8} is animated in the supplementary movie \textbf{2} that covers the time interval $0 \le t \le t_{sim}$.
The distributions resemble qualitatively those at $t=t_4$ and a stable state has been reached. The thickness of the interval, from which the protons were evacuated, has increased to about 0.5 and this is thus an ongoing process. The front of the cloud electrons has propagated from $x\approx 9.5$ at $t=t_4$ to about $x\approx 10.5$ at $t=t_{sim}$ at the speed $0.15c$. The high-density band in the protons and ambient electrons is accompanied by a thin electromagnetic sheath in Figs. \ref{fig8}(e,f). It does not fully enclose the pair plasma at the cloud front at $x\approx 10$ and we continue to observe an outflow of positrons at large $x$ and $|y| \le 2$. 

Density filaments, which are spatially correlated and approximately aligned with $x$, are present the distributions of the protons and ambient electrons in the interval $x>10$ and $|y|\le 1.5$. We can not observe magnetic field structures in Figs. \ref{fig8}(f) that follow these density striations. Their field amplitude is below the threshold $0.2$ and small compared to the magnetic fields driven by the clumpy positron distribution in this interval. 

The pair cloud is progressively separating itself from the ambient plasma close to the high-density band. A pair flow, which is separated from the ambient plasma, is a jet and hence we are observing its formation. The work that is required to expel the ambient plasma slows down the pair cloud close to the thin strong electromagnetic sheath and an inner cocoon forms. The accompanying movie shows structures in the density distribution of the pair cloud that are deflected sideways and slowed down at the thin strong electromagnetic sheath. The cloud electrons are separated from the ambient electrons at the head of the jet at $x\approx 8$ and the ambient protons are deflected sideways at $|y| \ge 2$. Our collisionless jet shows some features of a hydrodynamic jet.

\subsection{The electromagnetic piston and the pair flow at the head}

We investigate here the high-density band in the ambient plasma and the structure of the electromagnetic fields that sustain it. We focus on a jet interval where the high-density band is propagating orthogonally to the jet's main axis. 

\begin{figure*}
\includegraphics[width=\textwidth]{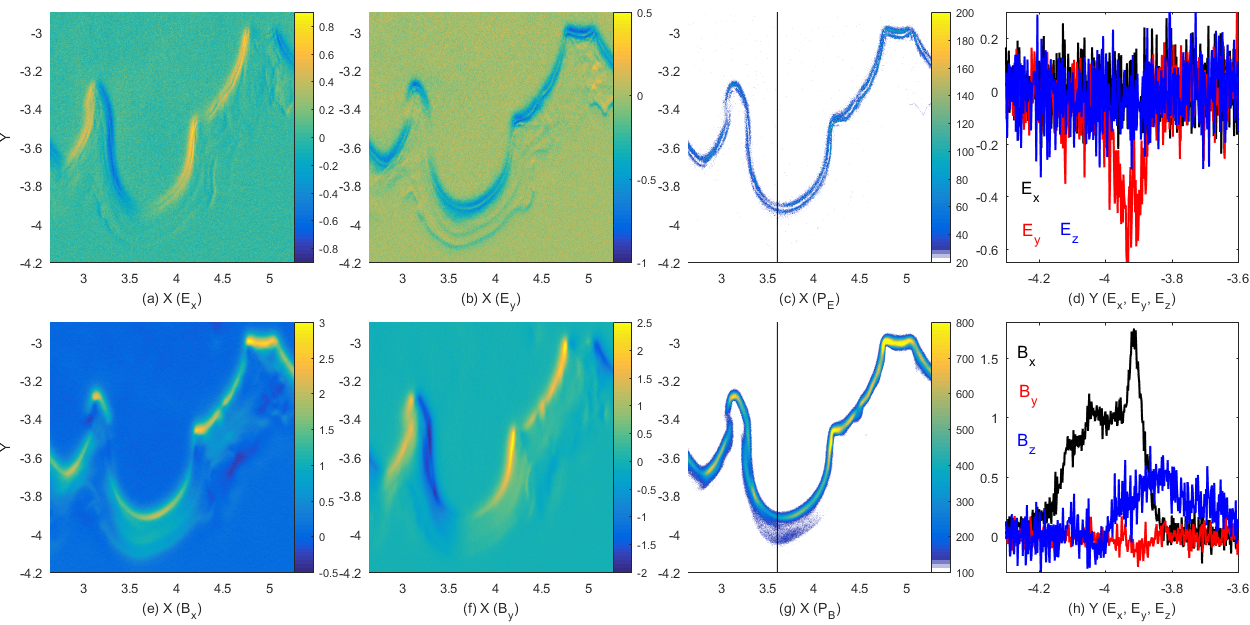}
\caption{Electromagnetic fields at the time $t=t_{sim}$ in full spatial resolution: panels (a) and (b) show the electric $E_x$ and $E_y$ components. The pressure $P_E = \epsilon_0(E_x^2+E_y^2)/(2P_{TE})$ is shown in panel (c). Panel (d) shows the electric field components along the vertical line in (c). Panels (e) and (f) depict the magnetic $B_x$ and $B_y$ components. Panel (g) shows the magnetic pressure $P_B=(B_x^2+B_y^2)/(\mu_0P_{TE})$ (color scale clamped to 800). Panel (h) shows the magnetic field components along the vertical line in (g). The vertical lines in (c, g) mark $x=3.6$.}
\label{fig9}
\end{figure*}

Figure \ref{fig9} shows the spatial distributions of the in-plane electric field components $E_x$ and $E_y$, those of the in-plane magnetic field components $B_x$ and $B_y$ and of their normalized energy densities $P_E = \epsilon_0(E_x^2+E_y^2)/2P_{TE}$ and $P_B = (B_x^2+B_y^2)/2\mu_0P_{TE}$ at the time $t=t_{sim}$. Lineouts of all field components are also plotted.
Both electric field components in Fig. \ref{fig9}(a, b) and their energy density show a banded structure, which is double-peaked  in some intervals. The electric field band follows the unipolar magnetic field band in Fig. \ref{fig9}(g). The amplitude of $B_x$ reaches more than 30 times the value $B_0=0.0884$ in this band. Large amplitudes of $B_y$ are observed Fig. \ref{fig9}(f) in the intervals where $B_x$ is weak and aligned with $y$ in Fig. \ref{fig9}(e); both components belong to the same band and their respective amplitudes depend on the band's orientation. Figure \ref{fig9}(g) reveals that the magnetic energy density in the center of this magnetic band exceeds $P_{TE}$ by a factor $\ge$ 200 almost everywhere. The energy density rises to more than 800 $P_{TE}$ in some intervals with a high curvature. We observe a thickening of the magnetic band for $3.4 \le x \le 4.1$. 

A comparison of the lineouts at $x=3.6$ in Figs. \ref{fig9}(d) and (h) shows that the large peak of $B_x$ at $x\approx -3.9$ coincides with a negative $E_y$ and that the other field components are weak at this location. We refer to this magnetic band and the electric field structure that is enclosing it as the electromagnetic piston. Figure \ref{fig9}(h) demonstrates that the amplitude of $B_z$, which is at least partially driven by the Weibel instability between the pair cloud and the protons behind the electromagnetic piston, is significantly lower than that of $B_x$ at $y=-3.9$. The mean value of $B_x$ over the interval $-4.4 \le y \le -4.3$ along $x=3.6$ is comparable to $B_0$ while the mean value in the interval  $-3.75 \le y \le -3.65$ is about $B_0/25$; the pair plasma has swiped out the background magnetic field and piled it up at the electromagnetic piston. 

\begin{figure}
\includegraphics[width=\columnwidth]{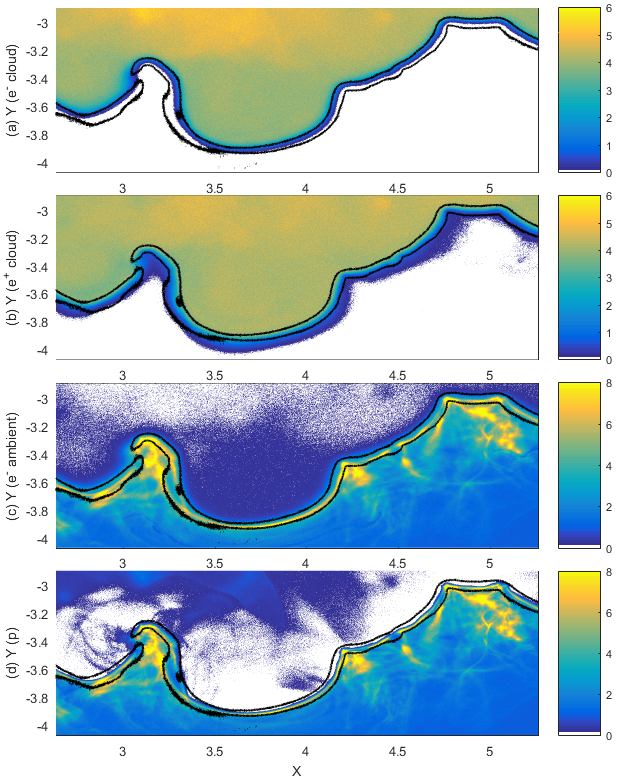}
\caption{Density distributions of the cloud electrons (a), of the positrons (b), of the ambient electrons (c) and protons (d) at the time $t=t_{sim}$. The contours $P_B=200$ are overplotted.}\label{fig10}
\end{figure}

We estimated the thermal gyroradius of an electron or positron of the cloud in a field with the amplitude $B_0$ as $r_{ge}\approx 0.54$. The magnetic field reaches a peak value of more than $15B_0$, which decreases the local thermal gyroradius to a value that is below the thickness of the electromagnetic piston. Figure \ref{fig10} depicts the effect the electromagnetic piston has on the particle populations.
The electrons and positrons of the cloud in Figs. \ref{fig10}(a,b) are indeed confined by the electromagnetic piston. The density distributions of the cloud particles behind the electromagnetic piston are almost uniform. Some positrons penetrate deeper into the electromagnetic piston than the electrons at $x\approx 3.6$ and $y\approx -3.9$ and their current may be responsible for the thickening of the magnetic band in Fig. \ref{fig9}(g).

Most ambient electrons in Fig. \ref{fig10} are confined to values of $y$ below those of the electromagnetic piston. They can not overcome the magnetic field of the electromagnetic piston and are pushed by it to lower $y$. Only few ambient electrons are found at larger $y$. The density of the ambient electrons exceeds 6 in intervals close to the cusps of the electromagnetic piston. 

The ambient protons are also mostly confined to values of $y$ that are lower than that of the electromagnetic piston. Even the strong magnetic field of the electromagnetic piston can not expel the protons via Larmor rotation. Its electric field is responsible for the proton acceleration. The proton density follows that of the ambient electrons and its peak value exceeds 6 close to the cusps of the electromagnetic piston. Some protons managed to break through the electromagnetic piston at $x=$ 3.2 and $y=$ -3.4. The magnetic field amplitude and, hence, the magnetic pressure of the electromagnetic piston are reduced by about 30\% at this location. The value of the magnetic pressure is below that of the contour line, which results in the apparent gap (see also Fig. \ref{fig9}(f)). A dilute proton population is found at large $y$ far behind the expanding electromagnetic piston. These protons were located behind the electromagnetic piston when it formed and hence they could not get swept out by it. 

\begin{figure}
\includegraphics[width=\columnwidth]{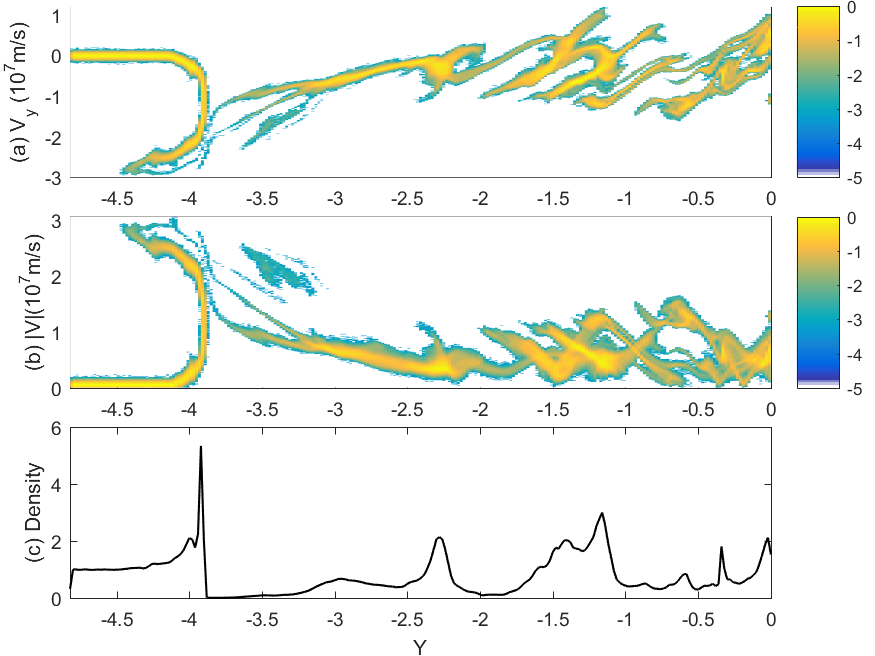}
\caption{The proton distribution along the slice $x=3.6$ for $y\le 0$ at $t=t_{sim}$. Panel (a) shows the phase space density distribution along the cut plane $y$ and $v_y$. Panel (b) shows the distribution along $y$ and the velocity modulus $|v|={(v_x^2+v_y^2+v_z^2)}^{1/2}$. The color scale is 10-logarithmic and normalized to the peak value of the distribution in each panel. The proton density distribution is shown in panel (c).}
\label{fig11}
\end{figure}

Figure \ref{fig11} shows the proton velocity and density distributions along the slice $x=3.6$.  
A single population of cool protons at rest, which retained their initial conditions, is located in the interval $y\le -4.5$. Protons are accelerated to a speed $-2.5 \times 10^7$ m/s at the position $y\approx -3.9$, which is where the electromagnetic piston is located in Fig. \ref{fig9}. Protons with the speed $2.5\times 10^7$ m/s have a kinetic energy of 4.7 MeV, which exceeds the thermal energy of the jet by an order of magnitude. Figure \ref{fig11}(b) demonstrates that the distribution of $y,v_y$ matches that of $y,|v|$ apart from the opposite sign. Protons in the lineout $x=3.6$ are thus accelerated only along $y$. If we assume that these protons have been reflected specularly by the electromagnetic piston then the latter is propagating at the speed $v_s = -1.25 \times{10}^7$ m/s or $-0.04c$. This speed matches the expansion speed we estimated by comparing the piston's locations along $y$ in Figs \ref{fig6} to \ref{fig8}. Figure \ref{fig11}(c) evidences that the proton density is practically zero behind the electromagnetic piston and below 0.1 for $-3.9 \le y \le -3.3$. 

The electromagnetic piston separates the jet plasma from the ambient plasma and its role is that of the contact discontinuity in hydrodynamic jets. Eventually a collisionless shock (shock 2 in Fig. \ref{fig1}) must form between the pristine ambient plasma, which is located at $y\le -4.5$ in Fig. \ref{fig11}(a), and the accelerated one in the interval $-4.5 \le y \le -3.9$. The nature of the shock will depend on how the collision speed between both ambient plasma populations compares to the characteristic plasma speeds. 

The ion acoustic speed in the ambient plasma is $c_s = {(k_B (\gamma_e T_e + \gamma_iT_i)/m_p)}^{1/2}$, where the electron and proton temperatures $T_e=T_p=T_0$. Electrons have 3 degrees of freedom $(\gamma_e = 5/3)$ in a collisionless plasma and protons 1 ($\gamma_p=3$) and $c_s \approx 10^6$ m/s. The electromagnetic piston has the speed $\approx 13c_s$. Electrostatic shocks, which are mediated by the ambipolar electric field across a density gradient, can not form at this high collision speed \citep{Forslund1971,Dieckmann2013} and the shock must be magnetized. 

The electromagnetic piston moves orthogonally across the initial magnetic field in Fig. \ref{fig11} and a shock will involve the fast magnetosonic mode. The Alfv\'en speed $v_A=B_0/(\mu_0 n_0 m_p)\approx 6.2\times 10^5$ m/s gives the fast magnetosonic speed $v_{fms}={(c_s^2+v_A^2)}^{1/2}\approx 1.1 \times 10^6$ m/s. The electromagnetic piston moves at the speed 11 $v_{fms}$ and the shock will be a supercritical fast magnetosonic shock \citep{Marshall1955}. Such shocks form on time scales that exceed an ion gyroperiod in the magnetic field $B_0$ \citep{Shimada2000,Chapman2005,Caprioli2014,Lembege2018}. It exceeds $t_{sim}$ by at least an order of magnitude. We can thus not observe yet the formation of an external shock between the pristine ambient plasma and the accelerated ambient plasma, which constitutes the outer cocoon. 

Figures \ref{fig11}(a,b) show energetic protons in the interval $y>-3.9$. The modulus of their peak speed is below $|v_s|$ and they are outrun by the electromagnetic piston. The distributions along $v_y$ and along $|V|$ hardly differ apart from the wrap-around at $v_y=0$ and they have thus also been accelerated mainly along $y$. Their peak speed $10^7$ m/s is comparable to that of the protons in the simulation by \citet{Dieckmann2018b} and they have thus been accelerated by the filamentation instability that formed initially between the pair cloud and the ambient plasma.

\begin{figure}
\includegraphics[width=\columnwidth]{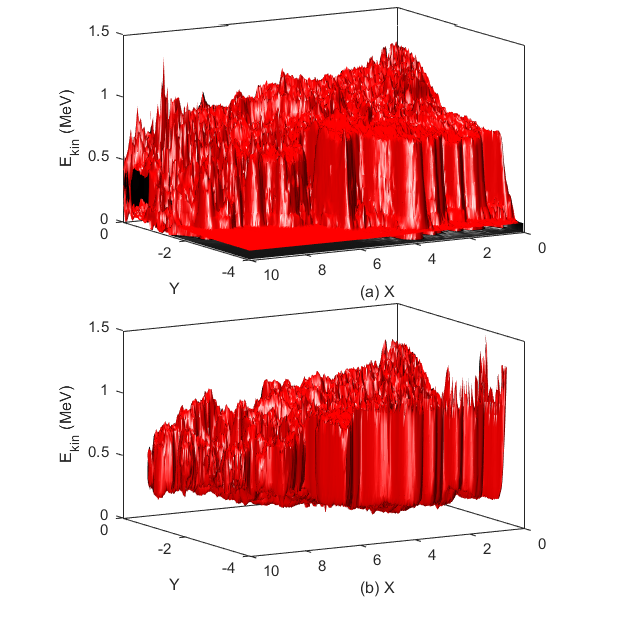}
\caption{The isosurfaces $30^{-1}$ for the electrons phase space density distribution $f_e(x,y,E_{kin})$ (a) and that of the positrons (b) at the time $t_{sim}$. The phase space densities are normalized to the peak density of the electrons.}
\label{fig12}
\end{figure}

\begin{figure}
\includegraphics[width=\columnwidth]{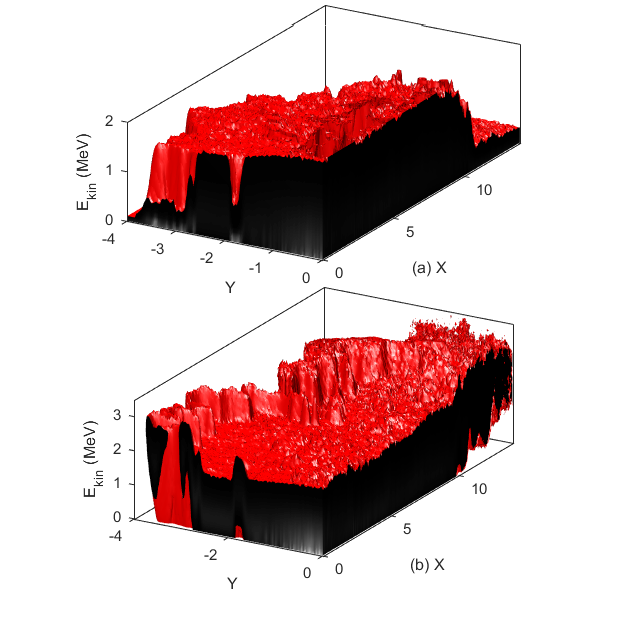}
\caption{The isosurfaces $30^{-2}$ for the electrons phase space density distribution $f_e(x,y,E_{kin})$ (a) and that of the positrons (b) at the time $t_{sim}$. The phase space densities are normalized to the peak density of the electrons.}
\label{fig13}
\end{figure}

\begin{figure}
\includegraphics[width=\columnwidth]{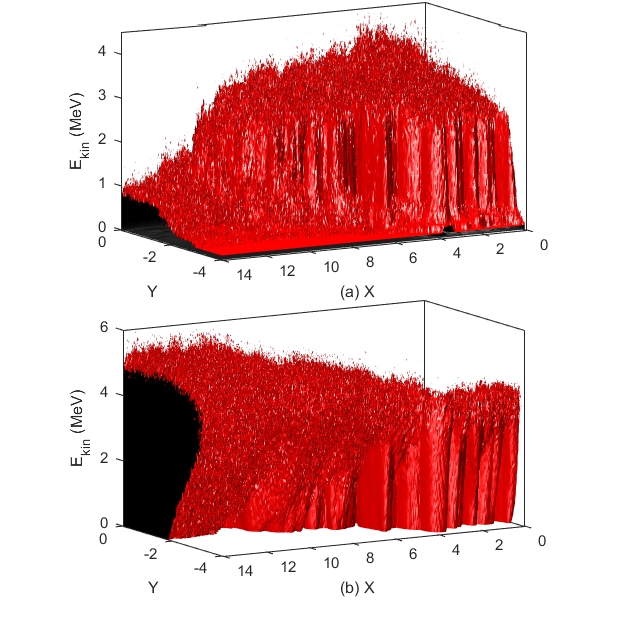}
\caption{The isosurfaces $30^{-3}$ for the electrons phase space density distribution $f_e(x,y,E_{kin})$ (a) and that of the positrons (b) at the time $t_{sim}$. The phase space densities are normalized to the peak density of the electrons.}
\label{fig14}
\end{figure}

We analyse the phase space density distributions of the positrons and electrons where we do not distinguish between ambient and jet electrons. We consider the phase space density as a function of $x,y$ and the energy $E_{kin} = \mathbf{p}^2/2m_e$, where $\mathbf{p}$ is the relativistic particle momentum. Phase space densities are integrated over 20 cells along x and along y and normalized to the peak density of the electrons. We show 3 isosurfaces that correspond to contours of the phase space density. The iso-surface with the density contour $30^{-1}$ in Fig. \ref{fig12} encloses the bulk of the particles.The density contour $30^{-2}$ in Fig. \ref{fig13} enwraps the particles with intermediate energies while the high-energy particles (density contour $30^{-3}$) are shown in Fig. \ref{fig14}. 

Figure \ref{fig12} shows the distributions of electrons and positrons with a low energy. 
Both distributions resemble each other for $x<8$ and for energies above 300 keV. Electrons with lower energies stem from the ambient electrons, which have no positronic counterpart in Fig. \ref{fig12}(b). Electrons and positrons are confined at large $|y|$. The bulk of the pairs is limited to energies below 1 MeV, which corresponds to the typical energy expected from leptons moving in a hot jet plasma with the mean speed 0.9c.

We observe significant differences between the populations of electrons and positrons in Fig. \ref{fig13}.
The electron distribution within the jet reaches a uniform maximum energy of 1.5 MeV for the selected density contour and it decreases rapidly at the jet boundary. We find more positrons with a higher energy close to the electromagnetic piston than inside the jet. 

Figure \ref{fig14} shows that the electrons and positrons reach comparable peak energies far from the electromagnetic piston and the jet's head at $x\approx 12$. More energetic positrons are observed close to the electromagnetic piston at $y\approx -3.5$ and $x<10$. 
Figure \ref{fig14} also reveals an outflow of energetic positrons along the jet spine $y \ge -3$, which has no counterpart in the electron distribution. The jet is thus a source of multi-MeV positrons that propagate along its expansion direction. 

Figures \ref{fig10}, \ref{fig13} and \ref{fig14} hint at how the electromagnetic piston is sustained. It forms a barrier for the ambient electrons (see Fig. \ref{fig10}(c)). As the electromagnetic piston swipes out the ambient electrons an electric field must grow that drags the protons with them (see Fig. \ref{fig10}(d)) to maintain the plasma's quasi-neutrality. The magnetic field of the electromagnetic piston can only be supported by the currents of relativistically fast particles. These currents must flow orthogonally to the simulation box in order to support the piston's in-plane magnetic field. Figures \ref{fig13} and \ref{fig14} revealed that we find more positrons than electrons with energies above 2 MeV close to the piston. The gyroradius 0.062 of an electron or positron with 2 MeV in a field with the magnetic amplitude 1.7 in Fig. \ref{fig9}(h) is comparable to the thickness of the electromagnetic piston. These energetic particles can thus penetrate deep into this structure, which is also demonstrated by Figs. \ref{fig10}(a, b). Having more energetic positrons than electrons implies that the electromagnetic piston is immersed in a net positronic current contribution. Figures \ref{fig5}-\ref{fig8} demonstrate that the electromagnetic piston is stable on time scales $t_{sim}$. The interplay of the electromagnetic piston with the energetic cloud particles around it yields a stable magnetic field configuration in the same way as magnetic boundaries in electron-ion plasmas can be sustained by the electron current \citep{Grad1961}.

\begin{figure}
\includegraphics[width=\columnwidth]{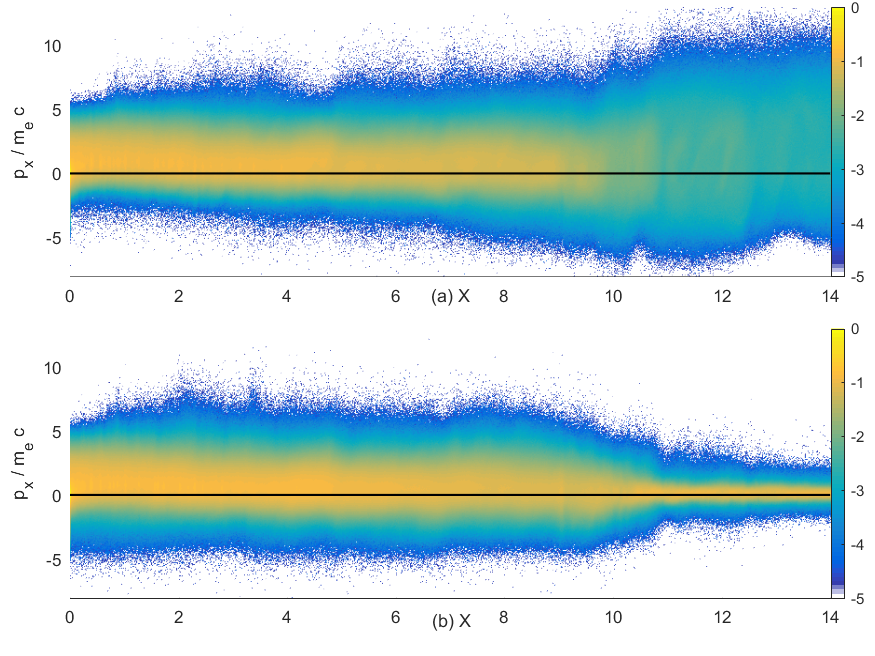}
\caption{The 10-logarithmic phase space density along $x$ and $p_x$ of the positrons (a) and electrons (b) in the interval $-0.35 \le y\le 0$. Both are normalized to the peak density in (b). Horizontal lines $p_x = 0$ are overplotted.}
\label{fig15}
\end{figure}

Figure \ref{fig15} shows the momentum distributions of the electrons and positrons along $x$ averaged over the interval $-0.35 \le y \le 0$.
Both distributions demonstrate that most electrons and positrons maintain their initial momentum in the interval $x \le 8$. A high-speed flow channel thus exists close to the spine of the collisionless jet. The momentum spread of the electrons decreases in the jet's head with $8 \le x \le 11$ and that of the positrons increases. We find more electrons with $p_x /m_e c\le -3$ in the interval $x \le 7$ than positrons. Jet electrons are reflected at the head and return to the injection point while positrons are accelerated at the head and stream out to larger $x$. This loss of positrons reduces the number of returning positrons in the interval $x \le 7$. Some ambient electrons reach mildly relativistic speeds in the interval $x>11$. They have been heated by the filamentation instability between the positrons and the ambient plasma. This filamentation instability is also heating up the positrons. Positrons in the interval $x>11$ are scattered by the electromagnetic fields to values $p_x<0$, which increases their thermal spread.   

\section{Discussion}

We have examined with a PIC simulation the interaction between a hot and fast cloud of electrons and positrons with a cool ambient magnetized plasma, which was composed of electrons and protons. The magnetic field was aligned with the mean flow direction of the pair cloud. The magnetic field amplitude was selected such that the gyroradius of the pair cloud's particles remained small compared to the distance between the pair cloud and the boundaries along $y$. The magnetic pressure remained small compared to the thermal pressure of the pair cloud. 

Initially the pair cloud expanded almost freely at its initial mildly relativistic speed. Larmor rotation slowed down the lateral expansion of the pair cloud while a filamentation instability developed between the pair cloud and the ambient plasma. Its magnetic field was significantly stronger than the background magnetic field and this instability was thus similar to that in the unmagnetized plasma observed by \citet{Dieckmann2018b}. This filamentation instability heated the protons to MeV temperatures. The expanding pair cloud swiped out the background magnetic field and piled it up into an electromagnetic piston. Its magnetic field vector was oriented in the simulation plane and this structure was thus not generated by the filamentation instability. Filamentation instabilities separate current filaments in the simulation plane and they yield magnetic fields that point orthogonally to the simulation plane.

The electromagnetic piston was strong enough to separate the pair cloud from the ambient plasma and it thus acted as a collisionless counterpart of a hydrodynamic contact discontinuity. It separated the pair cloud with its high thermal pressure from the ambient plasma with a much lower thermal pressure and magnetic pressure. Consequently the electromagnetic piston propagated into the ambient plasma. It pushed out the ambient electrons and an electric field grew that dragged the protons with it. A progressive separation of the cloud plasma from the ambient plasma evidences a forming collisionless jet. 

The propagation speed of the electromagnetic piston amounted to 0.04c in the direction that was orthogonal to the mean flow direction of the pair cloud. This propagation speed exceeded by far the thermal speed of the ambient protons and we observed a fast beam of protons that were reflected by the electromagnetic piston. The speed of this beam amounted to 13 times the ion acoustic speed $c_s$ in the ambient plasma. Electrostatic shocks can not form at such a high speed \citep{Forslund1971}. The beam speed was 11 times the fast magnetosonic speed $v_{fms}$ in the ambient plasma. Only supercritical magnetized shocks can form at such speeds. Their formation time exceeds an inverse ion gyrofrequency and we stopped our simulation long before such a shock could form. 

The jet front expanded at 0.15c along the magnetic field, which resulted in a shape that was elongated along the flow direction of the pair cloud. A reduction of the temperature of the pair plasma and an increase of its mean speed and of the magnetic field amplitude should result in a larger ratio between the longitudinal and lateral expansion speed of the jet.

Our simulation showed that the interplay of the electric and magnetic fields with the pair cloud resulted in a larger number of energetic positrons close to the electromagnetic piston. The positrons could penetrate deeper into the electromagnetic piston than the electrons and, consequently, a net current developed close to it that stabilized its magnetic field. A boundary of a collisionless jet that is formed by a spatially homogeneous strong magnetic field and is immersed in hot positrons and electrons, should give rise to radio emissions.

The electrons and positrons that flowed along the spine of the pair cloud maintained their initial flow speed until they reached the head of the pair cloud. Electromagnetic processes close to the jet's head slowed down the electrons of the pair cloud and accelerated its positrons, which resulted in an outflow of hot multi-MeV positrons along the jet axis. Some of these MeV positrons will flow along the magnetic field and escape into the ambient plasma far from the jet. Such positrons could contribute to the galactic positron population \citep{Panther2018}.

Previous simulations related to pair jets in unmagnetized plasmas \citep{Nishikawa2016,Arxiv} showed that magnetic fields grow in response to the filamentation- and mushroom instabilities \citep{Alves2015}. The magnetic field of the filamentation instability could become strong enough to act as a contact discontinuity in the simulation by \citet{Arxiv}. The latter simulation stabilized the jet with the help of a rigid spine. Here we have shown that a jet can also be stabilized by a guiding field with a thermal pressure that is low compared to the thermal pressure of the jet. The compressed background magnetic field replaces in this case that of the filamentation instability as the one that acts as a contact discontinuity. 

Here we drove the jet with a pair cloud but the jet was not purely leptonic. We launched the jet in an electron-proton plasma and some protons were behind the electromagnetic piston when it formed. The electromagnetic piston was also not impenetrable to the ambient protons. Some overcame it at locations where the magnetic pressure of the electromagnetic piston was reduced and entered the inner cocoon. Protons were accelerated on short time scales and by the end of the simulation the jet had a baryon component with an energy of the order MeV. The observed rapid proton acceleration was tied to the large relative speed between the protons and the pair cloud. This has consequences even for astrophysical jets. \citet{Waisberg2018} reported the observation of photo-ionization along the jet SS433. Neutrals will not interact with the electromagnetic piston and they can enter the inner jet. A filamentation instability sets in once these neutrals are ionized in significant numbers and they will rapidly be accelerated to MeV energies and possibly beyond.

Future studies will examine how the jet, the electromagnetic piston and the ambient plasma react to an ion beam that is propagating with the pair cloud. \citet{Fender2000} suggest that it is unlikely that a relativistic jet is composed solely of electrons and ions. It is however plausible that the jet carries with it a significant fraction of relativistic ions. The energy carried by a relativistic ion population implies that more energetic processes may develop than the ones we observed here.

\begin{acknowledgements}
M. E. D. acknowledges financial support by a visiting fellowship from the Ecole Nationale Sup\'erieure de Lyon, Universit\'e de Lyon. DF and RW acknowledge support from the French National Program of High Energy (PNHE). The simulations were performed with the EPOCH code financed by the grant EP/P02212X/1 on resources provided by the French supercomputing facilities GENCI through the grant A0030406960.
\end{acknowledgements}





\begin{thebibliography}{99}
\bibitem[Alves et al.(2015)]{Alves2015} Alves E.~P., Grismayer T., Fonseca R.~A., \& Silva L.~O. 2015, Phys. Rev. E, 92, 021101
\bibitem[Amato \& Arons(2006)]{Amato2006} Amato E., \& Arons J. 2006, Astrophys. J., 653, 325
\bibitem[Arber et al.(2015)]{Arber2015} Arber, T.~D., Bennett, K., \& Brady, C. S. et al. 2015, Plasma Phys. Controll. Fusion, 57, 113001
\bibitem[Bret et al.(2010)]{Bret2010} Bret, A., Gremillet, L. \& Dieckmann, M.~E. 2010, Phys. Plasmas, 17, 120501
\bibitem[Bromberg et al.(2011)]{Bromberg2011} Bromberg O., Nakar E., \& Piran T., et al. 2011, Astrophys. J., 740, 100
\bibitem[Caprioli \& Spitkovsky(2014)]{Caprioli2014} Caprioli D., \& Spitkovsky A. 2014, Astrophys. J., 783, 91
\bibitem[Chang et al.(2008)]{Chang2008} Chang P., Spitkovsky A., \& Arons J. 2008, Astrophys. J., 674, 378
\bibitem[Chapman et al.(2005)]{Chapman2005} Chapman S.~C., Lee R.~E., \& Dendy R.~0. 2005, Space Sci. Rev., 121, 5
\bibitem[Dal Pino(2005)]{DalPino2005} Dal Pino, E.~M.~D. 2005, Adv. Space Res., 35, 908
\bibitem[Dhawan et al.(2000)]{Dhawan2000} Dhawan V., Mirabel I.~F., \& Rodriguez L.~F. 2000, Astrophys. J., 543, 373
\bibitem[Diaz-Trigo et al.(2013)]{Trigo2013} Diaz Trigo M., Miller-Jones J.~C.~A., Migliari S., Broderick J.~W. \& Tzioumis T. 2013, Nature, 504, 260
\bibitem[Dieckmann et al.(2013)]{Dieckmann2013} Dieckmann M.~E., Ahmed H., Sarri G., et al. 2013, Phys. Plasmas, 20, 042111
\bibitem[Dieckmann \& Bret(2018)]{Dieckmann2018} Dieckmann M. ~E., \& Bret A. 2018, Mon. Not. R. Astron. Soc., 473, 198
\bibitem[Dieckmann et al.(2018a)]{Dieckmann2018a} Dieckmann M. ~E., Alejo A., Sarri G., Folini D., \& Walder R. 2018a, Phys. Plasmas, 25, 064502
\bibitem[Dieckmann et al.(2018b)]{Dieckmann2018b} Dieckmann M.~E., Alejo A., \& Sarri G. 2018b, Phys. Plasmas, 25, 062122
\bibitem[Dieckmann et al.(2018c)]{Arxiv} Dieckmann M.~E., Sarri G., Folini D., Walder R., \& Borghesi M. 2018c, Phys. Plasmas, 25, 112903
\bibitem[Esirkepov(2001)]{Esirkepov2001} Esirkepov, T.~Zh. 2001, Computer Phys. Comm., 135, 144
\bibitem[Falcke \& Biermann(1996)]{Falcke1996} Falcke, H., \& Biermann, P.~L. 1996, Astron. Astrophys., 308, 321
\bibitem[Fender et al.(2004)]{Fender2004} Fender, R.~P., Belloni, T.~M., \& Gallo, E. 2004, Mon. Not. R. Astron. Soc., 355, 1105
\bibitem[Fender \& Gallo(2014)]{Fender2014} Fender R. \& Gallo E. 2014, Space Sci. Rev., 183, 323
\bibitem[Fender \& Pooley(2000)]{Fender2000} Fender R.~P. \& Pooley G.~G. 2000, Mon. Not. R. Astron. Soc., 318, L1 
\bibitem[Ferriere(2001)]{Ferriere2001} Ferriere, K.~M. 2001, Rev. Mod. Phys., 73, 1031
\bibitem[Forslund \& Freidberg(1971)]{Forslund1971} Forslund D. ~W., \& Freidberg J.~P. 1971, Phys. Rev. Lett. 27, 1189.
\bibitem[Grad(1961)]{Grad1961} Grad H. 1961, Phys. Fluids, 4, 1366
\bibitem[Hededal \& Nishikawa(2005)]{Hededal2005} Hededal C. ~B., \& Nishikawa, K.~I. 2005, Astrophys. J., 623, L89
\bibitem[Jean et al.(2009)]{Jean2009} Jean P., Gillard W., Marcowith A., \& K. Ferri\`ere 2009, Astron. Astrophys., 508, 1099
\bibitem[Kazimura et al.(1998)]{Kazimura1998} Kazimura, Y., Sakai, J.~I., Neubert, T. \& Bulanov, S.~V. 1998, Astrophys. J., 498, L183
\bibitem[Lembege \& Yang(2018)]{Lembege2018} Lembege B. \& Yang Z.~W. 2018, Astrophys. J., 860, 84
\bibitem[Malkov et al.(2016)]{Malkov2016} Malkov M.~A., Sagdeev R.~Z., \& Dudnikova G.~I., et al. 2016, Phys. Plasmas., 23, 043105
\bibitem[Margon et al.(1979)]{Margon1979} Margon B., Ford H. C., Grandi S. A., \& Stone R. P. S. 1979, Astrophys. J., 233, L63
\bibitem[Marshall(1955)]{Marshall1955} Marshall W 1955, Proc. R. Soc. London A, 233, 367
\bibitem[Massaglia et al.(1996)]{Massaglia1996} Massaglia S., Bodo G., \& Ferrari A. 1996, Astron. Astrophys., 307, 997
\bibitem[Migliari et al.(2002)]{Migliari2002} Migliari S., Fender R., \& Mendez M. 2002, Science, 297, 1673
\bibitem[Mirabel \& Rodriguez(1999)]{Mirabel1999} Mirabel, I.~F., \& Rodriguez, L.~F. 1999, Astrophys. J., 37, 409
\bibitem[Neilsen et al.(2014)]{Neilsen2014} Neilsen J., Coriat M., Fender R., Lee J.\~C., Ponti G., Tzioumis A. K., Edwards P.~G. \& Broderick J.~W. 2014, Astrophys., 784, L5
\bibitem[Nishikawa et al.(2016)]{Nishikawa2016} Nishikawa K.~I., Frederiksen J.~T., Nordlund A., et al. 2016, Astrophys. J., 820, 94
\bibitem[Nishikawa et al.(2017)]{Nishikawa2017} Nishikawa K.~I., Mizuno Y., Gomez J. ~L., et al. 2017, Galaxies, 5, 58
\bibitem[Panther(2018)]{Panther2018} Panther F. ~H. 2018, Galaxies, 6, 39
\bibitem[Plotnikov et al.(2018)]{Plotnikov2018} Plotnikov I., Grassi A., \& Grech M. 2018, Mon. Not. R. Astron. Soc., 477, 5238
\bibitem[Poutanen et al.(2007)]{Poutanen2007} Poutanen J., Lipunova G., Fabrika S., Butkevich A. G., \& Abolmasov P. 2007, Mon. Not. R. Astron. Soc., 377, 1187
\bibitem[Schamel(1986)]{Schamel1986} Schamel H. 1986, Phys. Rep., 140, 161
\bibitem[Shimada \& Hoshino(2000)]{Shimada2000} Shimada N. \& Hoshino M. 2000, Astrophys. J., 543, L67
\bibitem[Siegert et al.(2016)]{Siegert2016} Siegert T., Diehl R., Greiner J., et al. 2016, Nature, 531, 341 
\bibitem[Silva et al.(2003)]{Silva2003} Silva L.~O., Fonseca R.~A., Tonge J.~W., et al. 2003, Astrophys. J., 596, L121
\bibitem[Spitkovsky(2008)]{Spitkovsky2008} Spitkovsky A. 2008, Astrophys. J., 673, L39
\bibitem[Waisberg et al.(2018)]{Waisberg2018} Waisberg I., Dexter J., Olivier-Petrucci P., Dubus G. \& Perraut K. 2018, Arxiv e-prints [arXiv:1811.12564]
\bibitem[Weibel(1959)]{Weibel1959} Weibel, E.~S. 1959, Phys. Rev. Lett., 2, 83
\end{thebibliography}


\end{document}